\documentclass[reprint,superscriptaddress,amssymb,aps,twocolumn,graphicx]{revtex4}

\usepackage{graphicx}% Include figure files
\usepackage{dcolumn}% Align table columns on decimal point
\usepackage{bm}% bold math
\usepackage{hyperref}

\usepackage{amsmath} 
\usepackage{mathrsfs}
\usepackage[cmintegrals]{newtxmath}
\usepackage[mathlines]{lineno}

\begin{document}

\bibliographystyle{unsrt}
\preprint{APS/123-QED}

\title{Exceptional point enhanced optical gyroscope in mechanical $\mathcal{PT}$-symmetric system}

\author{Xuan Mao}
\author{Guo-qing Qin}
\author{Hao Zhang}
\author{Hong Yang}
\affiliation{State Key Laboratory of Low-Dimensional Quantum Physics, Department of Physics, Tsinghua University, Beijing 100084, China}
\affiliation{Frontier Science Center for Quantum Information, Beijing 100084, China}

\author{Min Wang}
\email{mwangphys@mail.tsinghua.edu.cn}
\affiliation{State Key Laboratory of Low-Dimensional Quantum Physics, Department of Physics, Tsinghua University, Beijing 100084, China}
\affiliation{Frontier Science Center for Quantum Information, Beijing 100084, China}
\affiliation{Beijing Academy of Quantum Information Sciences, Beijing 100193, China}

\author{Gui-lu Long}
\email{gllong@tsinghua.edu.cn}
\affiliation{State Key Laboratory of Low-Dimensional Quantum Physics, Department of Physics, Tsinghua University, Beijing 100084, China}
\affiliation{Frontier Science Center for Quantum Information, Beijing 100084, China}
\affiliation{Beijing Academy of Quantum Information Sciences, Beijing 100193, China}
\affiliation{Beijing National Research Center for Information Science and Technology, Beijing 100084, China}

\date{\today}

\begin{abstract}

As an important device for detecting rotation, high sensitivity gyroscope is required for practical applications. In recent years, exceptional point (EP) shows its potential in enhancing the sensitivity of sensing in optical cavity. Here we propose an EP enhanced optical gyroscope based on mechanical $\mathcal{PT}$-symmetric system in microcavity. By pumping the two optical modes with different colors, i.e. blue and red detuning, an effective mechanical $\mathcal{PT}$-symmetric system can be obtained and the system can be prepared in EP with appropriate parameters. Compared with the situation of diabolic point, EP can  enhance the sensitivity of gyroscope with more than one order of magnitude in the weak perturbation regime. The results show the gyroscope can be enhanced effectively by monitoring mechanical modes rather than optical modes. Our work provides a promising approach to design gyroscope with higher sensitivity in optical microcavity and has potential values in some fields including fundamental physic and precision measurement.
 
\end{abstract}

\maketitle

%\tableofcontents

\section{INTRODUCTION \label{introduction}}

Based on the Sagnac effect\cite{post1967sagnac, chow1985ring}, a gyroscope with high precision plays an essential role in precision measurement. Various systems are involved in gyroscope including optomechanical systems\cite{lai2019observation, davuluri2017gyroscope, li2017microresonator}, photon and matter-wave interferometers\cite{haine2016mean, dowling1998correlated} and solid spin systems\cite{kornack2005nuclear, jaskula2019cross, wood2017magnetic}. High-quality optical microcavities\cite{vahala2003optical} have been promising platforms to investigate both fundamental physics and applications\cite{jiang2017chaos, lu2015p, liu2019sensing, zhang2017far, ward2018nanoparticle, wang2018rapid, zhang2019fast, naweed2005induced, liu2018gain, liu2018optothermal, xu2019frequency} in virtue of the ability to enhance light-matter interaction in an ultra-small volume. Among early studies, people have demonstrated exceptional point(EP)\cite{peng2014parity, wiersig2014enhancing, chen2017exceptional, hodaei2017enhanced, li2019nonreciprocal, peng2016chiral, chang2014parity, yi2019non, lu2018optomechanically, wang2019mechanical, miri2019exceptional} associated with non-Hermitian Hamiltonian governing the dynamic of an open system such as parity-time ($\mathcal{PT}$) symmetric system and proposed its applications in quantum sensing\cite{chen2018parity, degen2017quantum}, quantum metrology\cite{liu2016metrology}, low-threshold lasers\cite{jing2014pt, feng2014single, zhang2018phonon}, and so on.

Recently, sensitivity enhancement based on EP was presented in many detection schemes such as nanoparticle detection\cite{qin2019brillouin, wiersig2014enhancing, chen2017exceptional}, mass sensor\cite{djorwe2019exceptional} and gyroscope\cite{lai2019observation, ren2017ultrasensitive} both theoretically\cite{qin2019brillouin, djorwe2019exceptional,ren2017ultrasensitive, wiersig2014enhancing} and experimentally\cite{lai2019observation, chen2017exceptional} owing to the complex square-root topological behavior near an EP. Besides, optomechanical systems\cite{aspelmeyer2014cavity, weis2010optomechanically, dong2012optomechanical, fiore2011storing, jiang2015chip, qin2020manipulation, kronwald2013optomechanically, safavi2011electromagnetically, wang2019characterization, xu2015mechanical, chen2019phononic, liao2016macroscopic, shen2016experimental} have been considered as one of the potential candidates for sensing\cite{fan2011optofluidic, long2019overcoming, yao2017graphene, itoh2015quantum, yu2016cavity, nishiguchi2019single}. It is natural to investigate the related mode response before and after introducing perturbation according to which mode frequency shifts or mode splitting it causes. However, the fact of the narrower effective linewidth of mechanical modes inspires people to enhance the detection resolution by orders of magnitude\cite{yu2016cavity} in optomechanical systems. The advantage of monitoring the response of the mechanical modes, rather than measuring the frequency shift or the frequency splitting of the optical modes, is the minimal response required to distinguish in the sensing schemes has been reinforced.

As a combination of the enhancement performance of EP and the superiority of narrower linewidth possessed by mechanical modes, here we theoretically propose an enhanced gyroscope scheme in effective mechanical $\mathcal{PT}$-symmetric system. Reaching the balance of gain and loss, the system features EP in its mechanical spectrum, which guarantees the giant enhancement near EP. The proposed scheme differs from the known gyroscope in ($\romannumeral1$) it operates at mechanical EP rather than EP in the optical spectrum. ($\romannumeral2$) the gain and loss are self-engineered while the cavities are driven. In the absence of the rotation, we manage the system in the state of EP. In the presence of the perturbation, the eigenvalues are driven apart from the EP. Compared with conventional schemes utilizing diabolic point, EP enhances the sensitivity of gyroscope with more than $20$ times when the rotation frequency keeps below $1$ Hz. Moreover, the results show that the gyroscope can be enhanced effectively by monitoring mechanical response rather than optical modes change. Our work offers an alternative approach to design higher sensitivity gyroscope in the optomechanical system and has potential values in various research fields including fundamental physics and precision measurement.

This article is organized as follows: In Sec.\ref{basic model}, we demonstrate the basic model and the dynamical equations. We study the $\mathcal{PT}$-symmetric system and realize EP in Sec.\ref{PT}. We show the enhanced performance of EP and the superiority of monitoring mechanical modes response in Sec.\ref{EP}. In Sec.\ref{analysis}, different transmission spectra are presented. Conclusion is given in Sec.\ref{conclusion}.

\begin{figure}
    \centering
    \includegraphics[width=\linewidth]{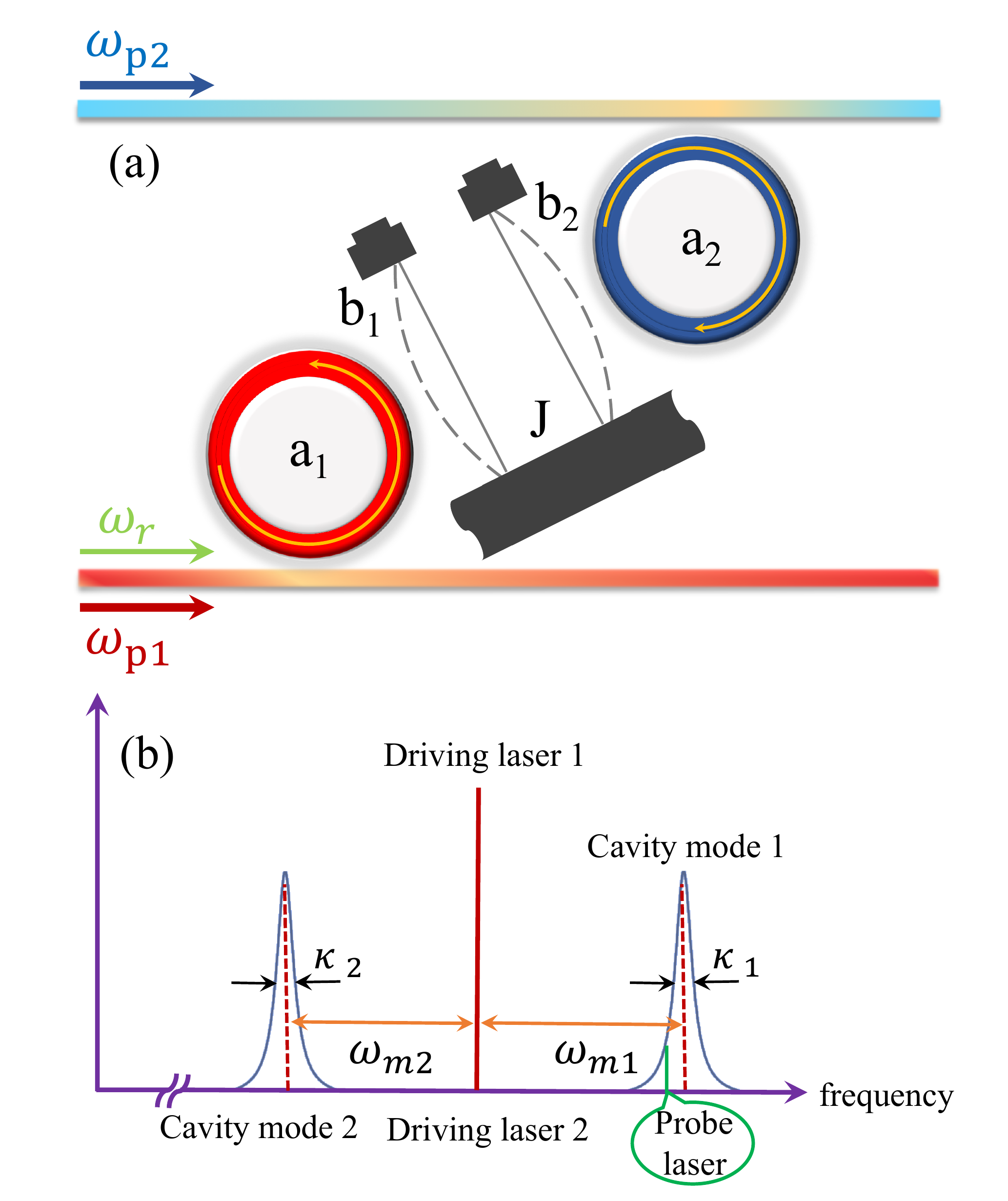}
    \caption{(a) Schematic of the mechanical $\mathcal{PT}$-symmetric system, which composed of two optical modes driven by red-detuning and blue-detuning pump laser respectively and two mechanical modes coupling to corresponding optical modes and coupling to each other simultaneously. (b) Frequency spectrogram of the generic optomechanical system. The frequency of the control lasers are degenerate while there is gap between the two cavity modes. The frequency difference between the driving laser 1 (driving laser 2) and cavity mode 1 (cavity mode 2) is $\omega_{m1}$ ($\omega_{m2}$).}
    \label{model}
\end{figure}

\section{MODEL AND DYNAMICAL EQUATIONS\label{basic model}}

The model we proposed is illustrated in Fig. \ref{model}  which contains two optical modes and two coupled mechanical modes. The first cavity is in the red-sideband regime while the other one is driven with a blue-detuned laser. The two optical modes couple to the corresponding mechanical mode respectively and the two mechanical modes couple to each other simultaneously. Owing to driving the two cavities symmetrically, it is feasible to manage either mechanical gain or loss. For non-rotation case, the Hamiltonian  describing the coupled optomechanical system is 

\begin{align}
    H &= H_{free} + H_{int} + H_{drive} + H_{probe}, \label{equation 1}
\end{align}

where ($\hbar = 1$)

\begin{align}
    H_{free} =& \omega_{a1} a_1^\dagger a_1 +\omega_{a2} a_2^\dagger a_2 + \omega_{m1} b_1^\dagger b_1 + \omega_{m2} b_2^\dagger b_2, \nonumber \\
    H_{int} =& g_1 a_1^\dagger a_1 (b_1^\dagger + b_1) + g_2 a_2^\dagger a_2 (b_2^\dagger + b_2) \nonumber\\
    &+ J (b_1^\dagger b_2 + b_1 b_2^\dagger) \nonumber\\
     H_{drive} =& i \sqrt{\kappa_{ex1}} \epsilon_{p1} e^{-i \omega_{p1} t} a_1^\dagger  + i \sqrt{\kappa_{ex2}} \epsilon_{p2} e^{-i \omega_{p2} t} a_2^\dagger + H.c. \nonumber\\
    H_{probe} =& i \sqrt{\kappa_{ex1}} S e^{-i \omega_r t} a_1^\dagger + H.c. , \label{equation 2}
\end{align}

 $H_{free}$ describes the free Hamiltonian of the optomechanical system, $a_i (b_i)$ and $a_i^\dagger (b_i^\dagger)$ (for $i=1,2$) are the annihilation and creation operators of the optical (mechanical) mode. The frequency and total decay of the optical modes here are $\omega_{ai}$ and $\kappa_i$, respectively. The mechanical resonators have the resonant frequency $\omega_{mi}$ with the effective mass $m$ and the decay rate $\gamma_i$. $H_{int}$ characterizes the interaction Hamiltonian of the model, the first two terms illustrate the cavity modes couple to the corresponding mechanical modes with single-photon optomechanical coupling strength $g_1$ and $g_2$ respectively. Also, there is mechanical-mechanical coupling with coupling rate $J$ as the third term of $H_{int}$ indicates. $H_{drive}$ implies the two optical modes are driven by external fields with strength $\epsilon_{pi}$ and frequency $\omega_{pi}$. $H_{probe}$ describes the probe laser characterized by strength $S$ and frequency $\omega_r$. In the rotating frame with the driving fields(to be simple, we consider the case of $\omega_{p1} = \omega_{p2} = \omega_p$) and after following the standard linearization procedure, the linearized equations of the fluctuation parts are expressed as 

\begin{align}
    \frac{\mathrm{d} a_1}{\mathrm{d} t} &= -i \Delta_1 a_1 - \frac{\kappa_1}{2} a_1 - i G_1 b_1 + \sqrt{\kappa_{ex1}} S e^{-i \delta t},  \label{equation 11}\\
    \frac{\mathrm{d} a_2}{\mathrm{d} t} &= -i \Delta_2 a_2 - \frac{\kappa_2}{2} a_2 - i G_2 b_2^\dagger,  \label{equation 12}\\
    \frac{\mathrm{d} b_1}{\mathrm{d} t} &= -i \omega_{m1} b_1 -\frac{\gamma_1}{2} b_1 - i G_1 a_1 - i J b_2,  \label{equation 13}\\
    \frac{\mathrm{d} b_2}{\mathrm{d} t} &= -i \omega_{m2} b_2 -\frac{\gamma_2}{2} b_2 - i G_2 a_2^\dagger - i J b_1. \label{equation 14}
\end{align}

Here, $\Delta_i = \omega_{ai} - \omega_p$ ($i=1,2$) represent the detuning between the optical mode and the driving field. $\delta = \omega_r - \omega_p$ is the detuning between the probe laser and the control field. $G_1$ and $G_2$ are the effective optomechanical coupling strength. If we focus on the transmission rate of the probe field, we can solve the equations above to get the steady-state solution of the fluctuation part of $a_1$, which is given below

\begin{gather}
    a_1 = - \frac{\sqrt{\kappa_{ex1}} S P}{P F + G_1^2}, \label{equation 15}
\end{gather}

Here, $P = -[A + {M J^2}/{(M B -G_2^2)}]$, $A = i (\omega_{m1} - \delta) + {\gamma_1}/{2}$, $B = i (\omega_{m2} - \delta) + {\gamma_2}/{2}$, $F = -[i (\Delta_1 - \delta) + {\kappa_1}/{2}]$ and $M = {\kappa_2}/{2} - i (\Delta_2 - \delta)$. According to the input-output formula, the output of the probe laser is expressed as $a_{out} = S - \sqrt{\kappa_{ex1}} a_1$. Without considering the high order sideband, the normalized transmission coefficient is $t = 1 - {\sqrt{\kappa_{ex1}} a_1}/{S}$ and the corresponding power transmission rate is $T = |t|^2$. On the other hand, under the assumptions of slowly varying amplitude and the evolution of mechanical modes are much slower than the optical modes ($G_i \ll \kappa_i$), we can derive the $\mathcal{PT}$-symmetric
dynamical equations of the coupled mechanical modes. 

\begin{align}
    \frac{\mathrm{d} b_1}{\mathrm{d} t} &= - (i \Omega_{m1} + \frac{\Gamma_1}{2} ) b_1 - i J b_2, \label{equation 21}\\
    \frac{\mathrm{d} b_2}{\mathrm{d} t} &= - (i \Omega_{m2} + \frac{\Gamma_2}{2} ) b_2 - i J b_1,  \label{equation 22}
\end{align}

where

\begin{align}
    \Gamma_1 &= \gamma_1+\frac{4 \kappa_1 G_1^{2}}{\kappa_1^{2}+4(\Delta_1-\omega_{m1})^{2}}, \label{equation 23}\\
    \Gamma_2 &= \gamma_2-\frac{4 \kappa_2 G_2^{2}}{\kappa_2^{2}+4(\omega_{m2}+\Delta_2)^{2}}, \label{equation 24}\\
    \Omega_{m1} &= \omega_{m1} - \frac{4 G_1^2 (\Delta_1 - \omega_{m1})}{\kappa_1^2 + 4(\Delta_1 - \omega_{m1})^2}, \label{equation 25}\\
    \Omega_{m2} &= \omega_{m2} - \frac{4 G_2^2 (\Delta_2 + \omega_{m2})}{\kappa_2^2 + 4(\Delta_2 + \omega_{m2})^2}. \label{equation 26}
\end{align}

Here $\Omega_{mi}$ and $\Gamma_i$ stand for the effective frequency and the effective decay of the $i$-th mechanical mode (for $i = 1,2$) respectively. And the effective Hamiltonian of the coupled mechanical resonators can be expressed as 

\begin{gather}
    H_{eff} = \left[
        \begin{array}{ccc}
             \Omega_{m1} - i \frac{\Gamma_1}{2}  & J \\
             J &  \Omega_{m2} - i \frac{\Gamma_2}{2} \\
        \end{array}
    \right]. \label{equation 27}
\end{gather}

The eigenvalues of the effective Hamiltonian $H_{eff}$ are given by

\begin{gather}
    \lambda_{\pm} = \omega_{\pm} + i \gamma_{\pm} = \frac{2(\Omega_{m1} + \Omega_{m2}) - i (\Gamma_1 + \Gamma_2) \pm \sqrt{\sigma}}{4}, \label{equation 28}
\end{gather}

where 

\begin{align}
    \sigma = & 16J^2 - (\Gamma_1 - \Gamma_2)^2 + 4(\Omega_{m1} - \Omega_{m2})^2 \nonumber\\ 
    &- 4i (\Omega_{m1} - \Omega_{m2})(\Gamma_1 - \Gamma_2). \label{equation 29}
\end{align}

The real parts $\omega_{\pm}$ and the imaginary parts $\gamma_{\pm}$ of the eigenvalues are corresponding to the effective frequency and linewidth of the mechanical modes, respectively. The exceptional point of the effective mechanical system occurs only if $\sigma = 0$. If $\Omega_{m1} = \Omega_{m2}$, the EP condition is equivalent to $4J = (\Gamma_1 - \Gamma_2)$. At this specific point, the two eigenvalues of the system are coalescing. After introducing rotation frequency $\Omega$ into the system, due to the Sagnac effect, the frequency of the optical modes experience a Sagnac-Fizeau shift \cite{malykin2000sagnac} which is shown as 

\begin{align}
    \omega_a &\rightarrow \omega_a \pm \Delta_{Sagnac}, \label{equation 30}\\
    \Delta_{Sagnac} &= \frac{n r \Omega \omega_{a}}{c} (1 - \frac{1}{n^2} - \frac{\lambda}{n} \frac{\mathrm{d} n}{\mathrm{d} \lambda}), \label{equation 31}
\end{align}

where $n$ and $r$ are the refractive index and the radius of the cavity and $c$ is the speed of light in vacuum. If the rotation speed $v = \Omega r$ is small enough compared to the light speed, the relativistic effects can be ignored, the dispersion term ${\mathrm{d} n}/{\mathrm{d} \lambda}$ (for typical material $\sim$ 1 $\%$ \cite{lv2017optomechanical}) becomes negligible. For our system, one can easily replace $\Delta_i$ to $\Delta_i \pm \Delta_{Sagnac}$ and the rest of caculation process is similar.

\section{MECHANICAL $\mathcal{PT}$-SYMMETRIC SYSTEM AND EXCEPTIONAL POINT \label{PT}}

\begin{figure}
    \centering
    \includegraphics[width=\linewidth]{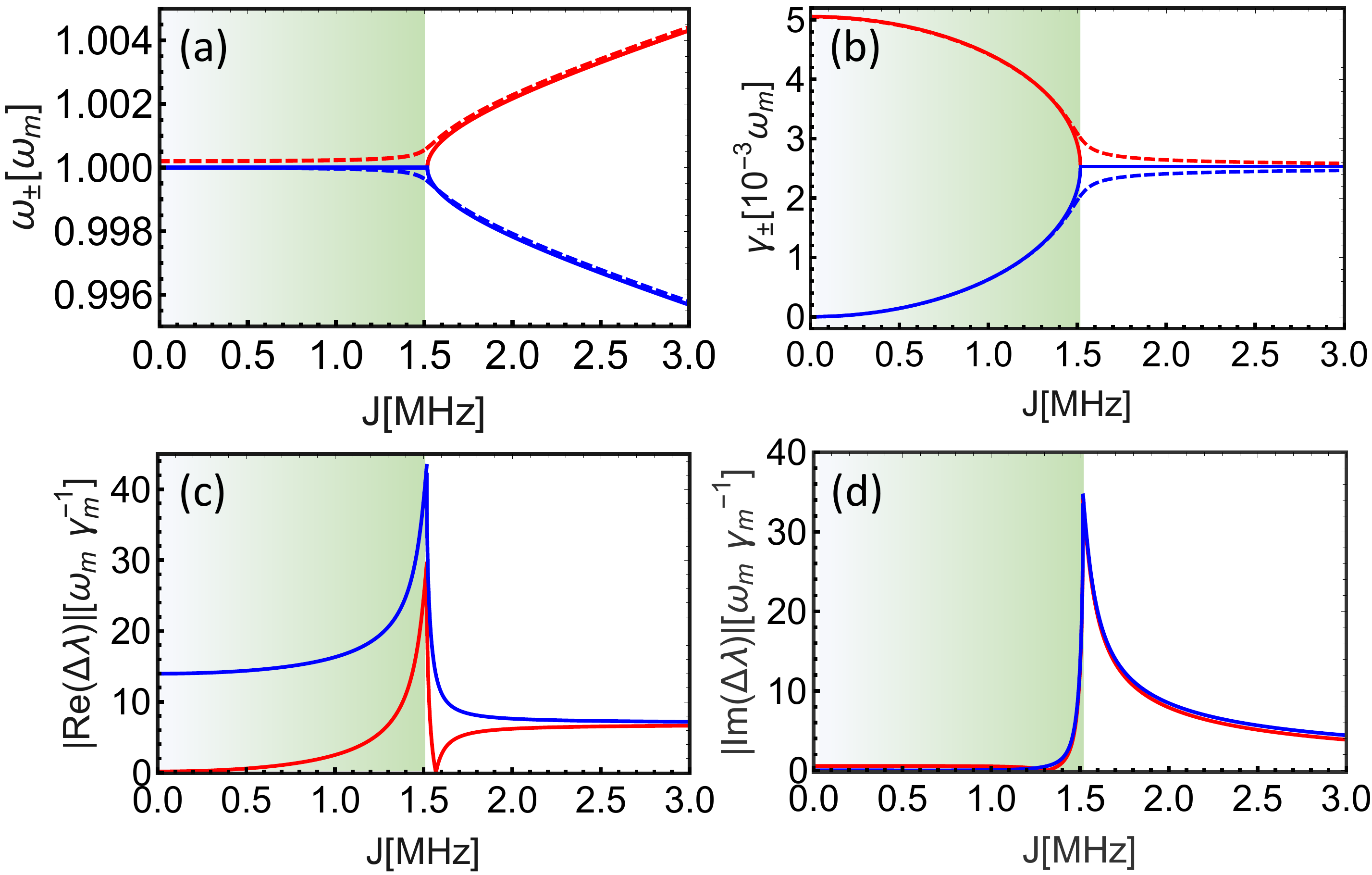}
    \caption{(a), (b) The real and imaginary parts of the eigenvalues versus the mechanical-mechanical coupling strength $J$ before (solid lines) and after (dashed lines) perturbation introduced by rotation with corresponding optical frequency shift $\Delta_{Sagnac} = 2*10^{-4} \omega_{m}$. The two eignvalues of the effective Hamiltonian are marked with different color lines. (c), (d) Gap difference between the solid lines and the dashed lines shown in (a) and (b). The $\mathcal{PT}$ -symmetric region and the $\mathcal{PT}$ -symmetric broken region are marked with different colors in this figure. All the quantities are plotted versus $J$ and note the difference of unit of y axis in (a) (b) and (c) (d). The other parameters used in this system are $\kappa_1 = \kappa_2 = 6$ MHz, $\omega_{m1} = \omega_{m2} = \omega_{m}$ = $600$ MHz, $\gamma_1 = \gamma_2 = 70$ kHz and $c_2 = {4G_2^2}/{(\kappa_2 \gamma_2)} = 0.9999$.}
    \label{perturbation}
\end{figure}

Firstly, we focus on the case of $\Omega = 0$. As Eq. (\ref{equation 27}) - Eq. (\ref{equation 29}) have shown, the optomechanical system proposed can be regarded as the effective mechanical $\mathcal{PT}$-symmetric system. And for some specific parameter values, the mechanical system can reach exceptional point (EP). The two eigenvalues of the system are coalescing at the specific point. Fig. \ref{perturbation} illustrates the real and imaginary parts of the eigenvalues versus the mechanical-mechanical coupling strength $J$ before (solid lines) and after (dashed lines) perturbation introduced by rotation with corresponding optical frequency shift $\Delta_{Sagnac} = 2*10^{-4} \omega_{m}$. The other parameters we used here are $\kappa_1 = \kappa_2 = 6$ MHz, $\omega_{m1} = \omega_{m2} = \omega_{m}$ = $600$ MHz, $\gamma_1 = \gamma_2 = 70$ kHz, $n = 1.44$, $r = 9$ mm, $\omega_a = 2*10^{14}$ Hz and $\lambda = 1550$ nm. Considering the experimental generality, we choose $\kappa_{exi} = \kappa_i / 2 $ (for $i = 1,2$), which indicates the coupling between the cavity and fiber taper is in the critical coupling zone. On account of the cavity mode $a_2$ is in the blue sideband, it is necessary to mention that the effective coupling strength $G_2$ can't be too strong. Here $G_2$ in our system satisfies $c_2 = {4G_2^2}/{(\kappa_2 \gamma_2)} = 0.9999$ to avoid unstable situation emerging. 

As Fig. \ref{perturbation} demonstrates, one can modulate the effective mechanical $\mathcal{PT}$-symmetric system transiting between $\mathcal{PT}$-symmetric region and $\mathcal{PT}$-symmetric broken region by modifying the system parameters such as the effective optomechanical coupling strength $G_1$ and mechanical-mechanical coupling strength $J$. In Fig. \ref{perturbation}, we have marked the $\mathcal{PT}$-symmetric region and $\mathcal{PT}$-symmetric broken region with different colors. In the $\mathcal{PT}$-symmetric region, the eigenvalues of the system share the same real part but exhibit different imaginary parts, which is to say the two eigenvalues of the system present the same frequency while possessing different linewidths. On the contrary, the eigenvalues have the same linewidth with disparate frequencies when the system is in the $\mathcal{PT}$-symmetric broken region. There is a transition point in the figure i.e. $J = 1.5175$ MHz. It is observed that there is EP feature in the effective mechanical system. 

For our effective mechanical system, the two eigenvalues coalesce at EP and any perturbation will induce the eigenvalues splitting. For any tradition gyroscope, the response of the system is proportional to the strength of the perturbation \cite{chen2017exceptional}. There is a completely different situation in our system. Taking the perturbation induced optical frequency shift $\Delta_{Sagnac} = 2*10^{-4} \omega_{m1}$ for example, Fig. \ref{perturbation} exhibits how the real parts and the imaginary parts of the eigenvalues change when we change the magnitude of $J$ before (solid lines) and after (dashed lines) the perturbation. As Eq. (\ref{equation 27}) has shown, the frequency and the linewidth of the mechanical modes are corresponding to the real parts $\omega_{\pm}$ and the imaginary parts $\gamma_{\pm}$, respectively. In Fig. \ref{perturbation} (a), the frequency of the mechanical modes transit from degenerate to non-degenerate after introducing the external perturbation at EP. Fig. \ref{perturbation} (c) indicates the gap difference between the solid lines and the dashed lines reaches its maximum value at EP for the same perturbation strength and the same conclusion can be inferred from Fig. \ref{perturbation} (d) for the linewidth of the mechanical modes. It should be noted that the frequency of mechanical modes move toward the  opposite directions, which solids the foundation of applying the EP into the gyroscope in our proposal. Another interesting feature of the EP is the damping of the mechanical modes are degenerate as depicted by solid lines in Fig. \ref{perturbation} (b). Introducing rotation into this system will also lift the degeneracy (see the difference between the solid lines and the dashed lines). Different from the gap of the real parts, the linewidth response magnitude of the mechanical modes are exactly the same while they change toward the opposite directions. It is related to the physics fact that one of the mechanical modes experiences gain whereas the other one experiences loss in this process. That provides the evidence that our proposal is truly a $\mathcal{PT}$-symmetric system and we can manage either mechanical gain and loss symmetrically. By further increasing the rotation angular frequency $\Omega$, the linewidth of the mechanical modes will be out of the vicinity of the EP. 

\begin{figure}
    \centering
    \includegraphics[width=\linewidth]{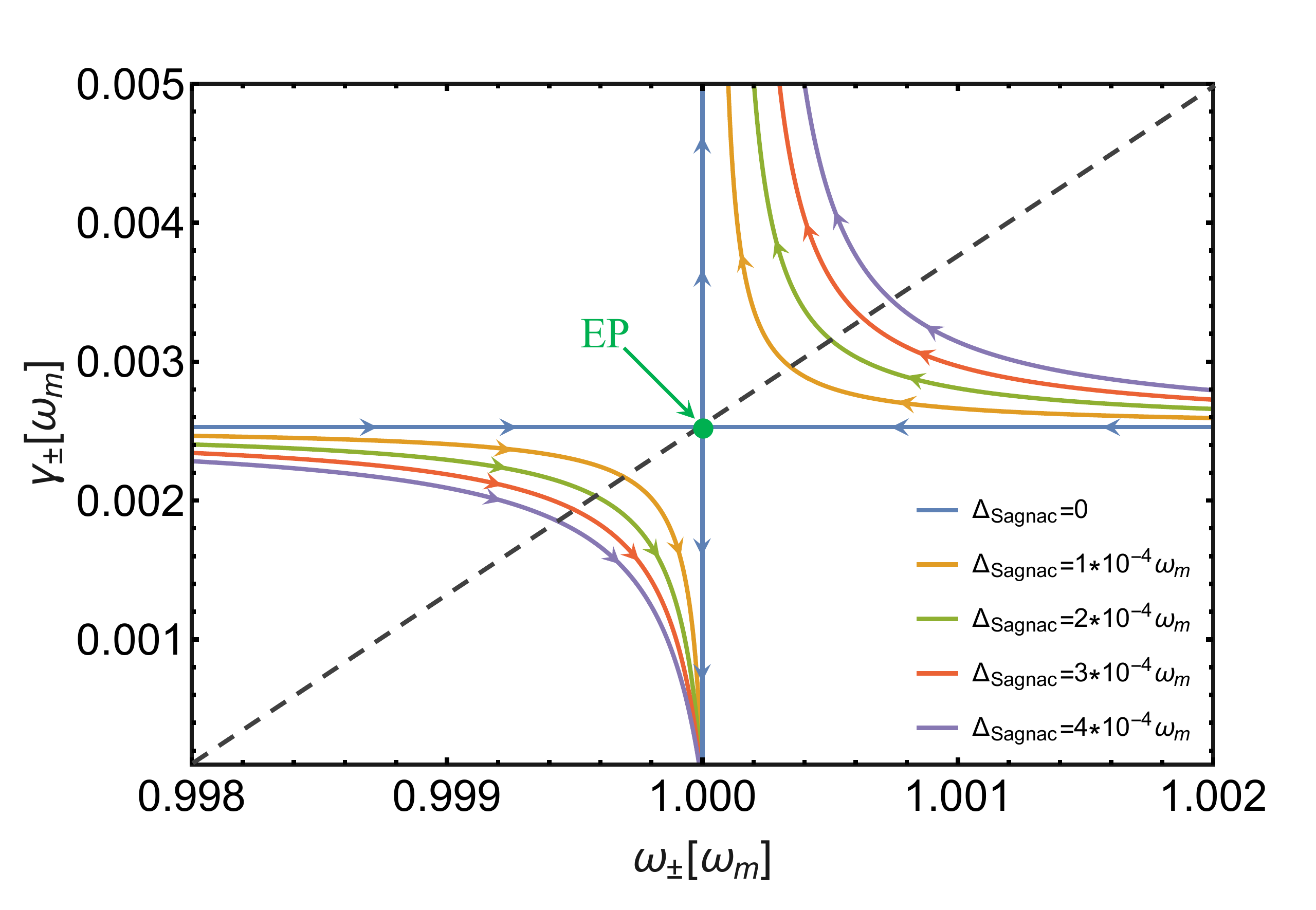}
    \caption{Trajectories of the eigenvalues in the perturbation plane for different rotation speed characterized by different $\Delta_{Sagnac}$. For the perturbed systems,  the
    distance between the intersections of the associated curves (depicted
    by the same color)  is bigger and bigger as the rotation speed increases. The arrows of the figure indicate the evolution of the eigenvalues as the mechanical-mechanical coupling strength $J$ increases. And the black dashed line implies $4J = (\Gamma_1 - \Gamma_2)$ condition. Note that the exceptional point is located at the center of the plane and has already marked by the green dot. The parameters are the same as given in Fig. \ref{perturbation}.}
    \label{complex}
\end{figure}

To characterize the effective mechanical system under different rotation speeds, the trajectories of the eigenvalues in the perturbation plane are plotted in Fig. \ref{complex}. Applying different perturbation strength into the effective mechanical system, the evolution behaviors of the eigenvalues are similar except the distance of the associated curves (implied in the same color). The arrows in this figure imply the direction of increasing $J$ and the EP of the system is located in the center of the plane (marked by the green dot). As the rotation speed grows, the distance between the same color curves becomes bigger and bigger. As expected, for the same perturbation strength (lines in the same color in the figure), the distance to the unperturbed system (shown by the blue lines) achieves maximum at $J = 1.5175$ MHz with the same parameter values. 

\section{ENHANCING SENSITIVITY AT EXCEPTIONAL POINT \label{EP}}

A natural idea of detection is observing the related mode response (usually the frequency shift or the frequency splitting) before and after introducing perturbation. If one intended to detect rotation speed or nanoparticles, the frequency shift or the linewidth change of the optical modes will be essential from this point of view. In parallel, the shift of the mechanical modes is a crucial quantity for experimentalists for mass or displacement sensing schemes. In our proposal, we observe the mechanical modes frequency shift under the rotation frame even though it only induces an optical frequency shift. The motivation of this thought is taking good advantage of the much narrower linewidth possessed by the mechanical modes. Different from EP, some special micro-cavities such as micro-toroidals support modes with degenerate eigenvalues but the associated eigenvectors can always be chosen to be orthogonal to each other, also known as DP. For traditional cavity-based gyroscope schemes, the DP is utilized and the frequency shift in response to the perturbation strength $\epsilon$ is proportional to $\epsilon$. In contrast, the response operating at EP in our system is proportional to $\epsilon^{1/2}$. To make full use of the difference between the DP and the EP, one can easily infer that the sensitivity is greatly enhanced by using the EP for sufficiently small perturbations. The enhanced effect is the result of the intrinsic properties of the EP and has nothing to do with the cavity type, the materials, or the perturbation properties. 

\begin{figure}
    \centering
    \includegraphics[width=\linewidth]{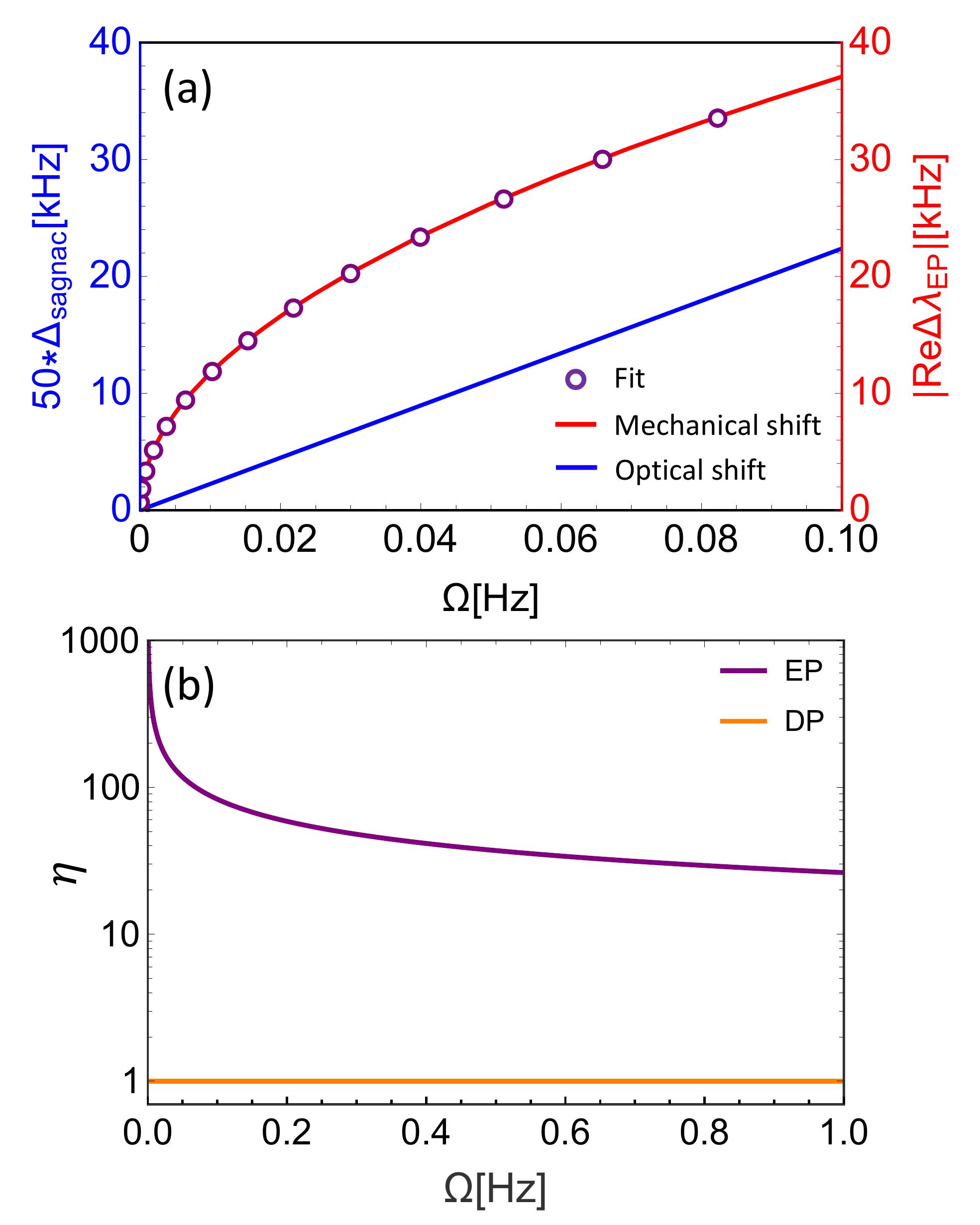}
    \caption{(a) The frequency shift of the mechanical mode and optical mode versus rotation frequency $\Omega$ near EP are illustrated by the red solid line and blue solid line respectively. The open circles mark the fit trend with $F = 2.13*10^7$. (b) The contrast of the EP and DP when the rotation frequency keeps below $1$ Hz. Note that the curves are plotted in a logarithmic graph. The parameters are the same as given in Fig. \ref{perturbation}.}
    \label{enhancement}
\end{figure}

As mentioned above, the frequency shift of the mechanical modes in our effective system, i.e. $|Re \Delta \lambda| = |Re\lambda_+ - Re\lambda_-|$, can be expressed as proportional to the square root of perturbation strength near the EP, which means we can fit the shift in response to the rotation angular frequency with

\begin{gather}
    |Re \Delta \lambda| = F \Omega ^ {1/2}. \label{equation 32}
\end{gather}

To investigate the superiority of utilizing EP and the mechanical modes intuitively, we compare the optical shift to the mechanical shift versus the rotation frequency $\Omega$ in Fig. \ref{enhancement} (a) and demonstrate the EP sensitivity and the traditional DP sensitivity in Fig. \ref{enhancement} (b). For the same $\Omega$, the mechanical shift (shown by the red line) is always more remarkable than the optical shift (implied by the blue line). It should be noted that the difference between the blue y-axis label and the red one. The red axis reveals the difference between the real part of the two eigenvalues at EP while the blue axis illustrates 50 times the optical shift at EP parameter values, which indicates the mechanical response is more than 50 times to the optical shift when the rotation frequency keeps below 0.1 Hz under the same parameter values. The performance of the mechanical modes is even better if the rotation frequency is sufficiently small as the red curve shows. This is in our expect owing to the properties of the EP in our system. The open circles suggest the fit trend as Eq. \ref{equation 32} expressed with $F = 2.13*10^7$, which has a great agreement with the mechanical shift in our proposal. One can safely infer that the mechanical shift in response to the external rotation frequency $\Omega$ obeys the square root behavior and that is the direct result of utilizing EP rather than DP in the scheme. 

The optomechanical system operating at EP exhibits the great enhancement factor for rotation sensing. The enhancement factor is used to describe how much better performance that the response of the sensor operating at EP in comparison to the response of the traditional sensor utilizing DP. The enhancement factor can be defined as 

\begin{gather}
    \eta \equiv |\frac{Re \Delta \lambda}{\Delta_{Sagnac}}|. \label{equation 33}
\end{gather}

For the conventional sensors utilizing DP rather than EP, at which the eigenvalues are degenerate while the corresponding eigenvalues are orthogonal, the response of the sensor is proportional to the perturbation strength and the enhancement factor is 1 all the time. But for the sensors based on EP, in the absence of the rotation, the system is near EP condition. In the presence of the perturbation, the eigenvalues of the effective mechanical system depart from the vicinity of the EP, which gives us the inspiration to detect the external perturbation operating at EP. Fig. \ref{enhancement} (b) illustrates that the EP sensitivity is much better than the DP sensitivity for the weak perturbation. The smaller the rotation frequency can reach, the better performance that the EP gyroscope can exhibit while the DP sensor maintains the same performance no matter what the rotation frequency is. That simply proves the efficiency of the sensors based on EP in detecting small nanoparticles, temperature variation, rotation frequency and so on. If the perturbation strength is large enough, the response of the EP sensor is no longer proportional to $\epsilon^{1/2}$ and will approach linear. As a result of large perturbation strength, the enhancement factor evolves toward the limit $\eta \sim 1$ and can be inferred from Fig. \ref{enhancement} (b). 

It should be noted that the effective linewidth of one of the mechanical modes will become narrower and the other one will be wider owing to pumping the two optical modes with different colors. The linewidth of the mechanical modes is depicted by Eq. \ref{equation 23} and Eq. \ref{equation 24} and under the parameter values shown in Fig. \ref{perturbation} one can get $\Gamma_1 = 6.07$ MHz and $\Gamma_2 = 7$ Hz. Because of the ultra-narrow effective linewidth, it is unnecessary to consider the minimum detectable signal in the limitation of the measurement device. 

Our proposed optomechanical system consisting of two optical modes and two mechanical modes has its own experimental value thanks to the progress of nanofabrication. Under the consideration of the feasibility of experiments, the parameter values in our proposed scheme are operable. Our scheme provides a new concept of using mechanical response rather than testing the optical shift directly operating at exceptional point theoretically. Moreover, the values of system parameters are chosen carefully under the consideration of  feasibility to bridge the theoretical proposal and experiments.

\section{ANALYSIS OF THE TRANSMISSION SPECTRA\label{analysis}}

\begin{figure}
    \centering
    \includegraphics[width=\linewidth]{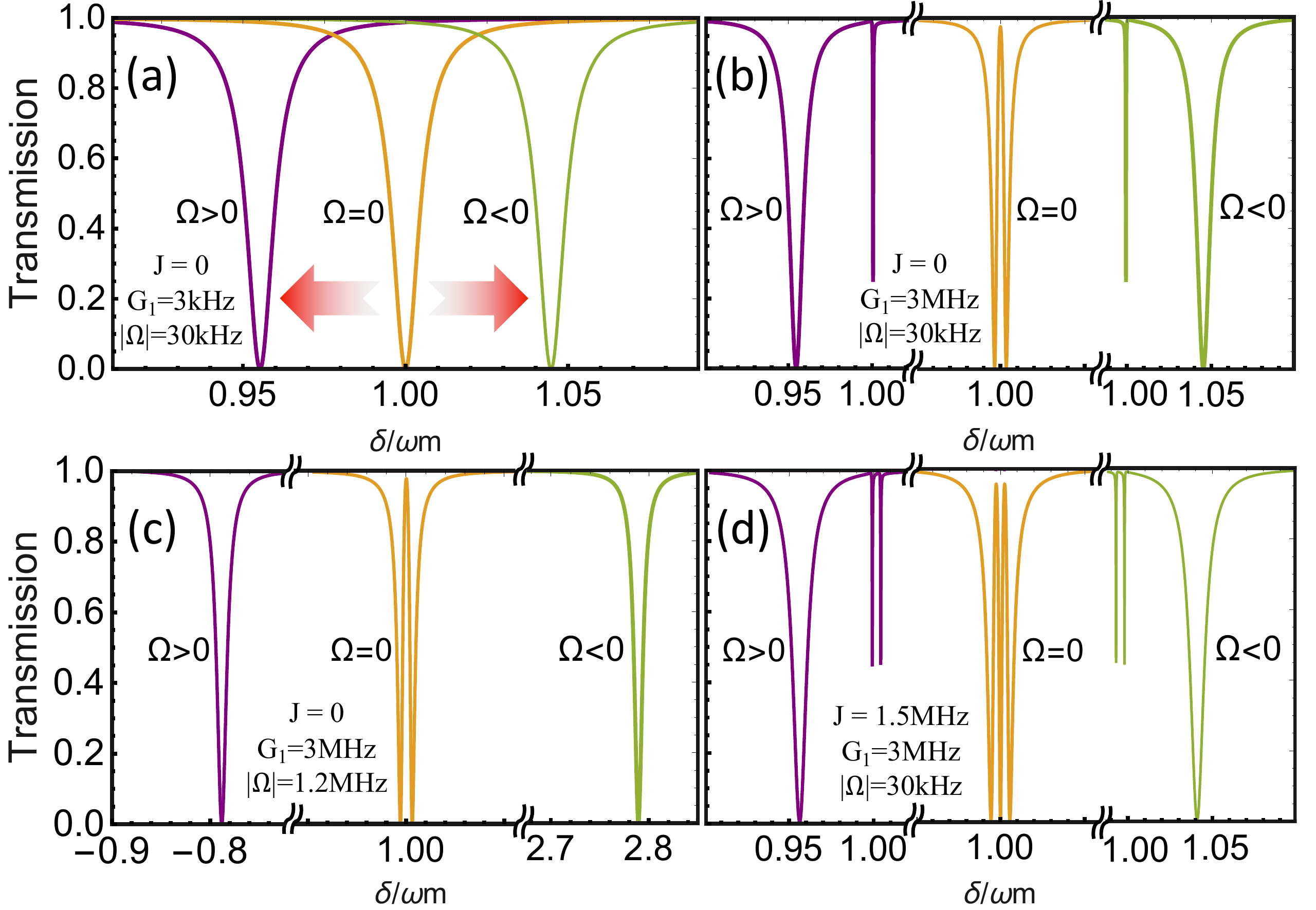}
    \caption{The transmission rate of the cavity mode under different parameter values. (a) The resonant frequency of the cavity mode can be modulated by both the magnitude of the rotation frequency $\Omega$ and the rotation direction. (b) For the case of $\Omega = 0$, OMIT occurs when the effective optomechanical coupling strength $G_1$ increases. (c) The OMIT phenomenon disappears when the  rotation frequency is sufficiently high. The insets illustrate that there is OMIT when $\Omega = 0$ while OMIT disappears when $\Omega = 1.2$ MHz. (d) There occurs a dip within the OMIT window when the mechanical-mechanical coupling strength $J$ reaches to $1.5175$ MHz. The other parameters are the same as given in Fig. \ref{perturbation}.}
    \label{transmission}
\end{figure}

From the perspective of experiments, the frequency shift that the mechanical modes experience can alter the optical responses. For our system demonstrated in Fig. \ref{model}, it is instructive to study the transmission spectra under different values of the mechanical-mechanical coupling strength $J$, the effective optomechanical coupling strength $G_1$ and the rotation frequency $\Omega$. Fig. \ref{transmission} illustrates the transmission rate $T$ varies with the detuning between the probe laser and the control field $\delta$ in the unit of mechanical frequency $\omega_{m}$. 

First we consider the simplest case $J = 0$ and the effective optomechanical coupling strength $G_1$ is sufficiently small. It is expected that the transmission rate presents the Lorenz curve as Fig. \ref{transmission} (a) shows. The resonant frequency of the cavity mode can be modulated by both the magnitude of the rotation frequency $\Omega$ and the rotation direction. If $\Omega > 0$, that means the rotation direction is consistent with the direction of the cavity mode. According to the Sagnac effect, the optical mode frequency is expected to decrease by an amount of $\Delta_{Sagnac}$. Thus the resonant frequency of the optical mode decrease. The larger the rotation speed is, the larger the amount of frequency shift experiences. On the contrary, if $\Omega < 0$, the resonant frequency associates with an increasing amount of $\Delta_{Sagnac}$. On the basis of the case (a), we enlarge the magnitude of $G_1$ to 3 MHz. Optomechanical induced transparency (OMIT) phenomenon turns out due to different pathways interference effect as implied by the orange line in Fig. \ref{transmission} (b). The first pathway is the probe photons excite the cavity mode $a_1$ and couple to the output port and the other one is the photons generated by the sideband transition through the optomechanical interaction in $a_1$ are coupled out of the cavity. In the case of rotation angular frequency is not very large, there are still two dips in the transmission spectrum. As the rotation frequency increases, the width of the transparency window becomes larger accompany with the depth of one of the dips diminishes. Note that the frequency shift of the other dip is exactly the same as shown in Fig. \ref{transmission} (a) in the condition of the same rotation speed. Furthermore, we  expand the rotation angular frequency $\Omega$ to 1.2 MHz, the OMIT vanishes illustrated in Fig. \ref{transmission} (c). There is OMIT when $\Omega = 0$ while the transmission spectrum back to the Lorenz curve in the case of $\Omega = 1.2$ MHz. The rotation frequency becomes sufficiently high, the frequency shift of the optical mode extend too large to neglect, leading to great frequency difference to destroy the destructive interference effect.

In the case of the mechanical-mechanical coupling strength $J$ reaches to 1.5175 MHz, there occurs a narrow dip within the transparency window demonstrated in Fig. \ref{transmission} (d). Similarly, as the rotation speed $\Omega$ increases, the transparency window becomes wider and wider and two of the dips in the spectrum become shallower and shallower. Meanwhile, the other dip maintains the same frequency shift under the condition of the same value of $\Omega$. We can safely infer that if the rotation frequency is extremely high the dip within the OMIT window and OMIT will vanish. It is worth pointing out that the response of the optical frequency becomes recognizable requires the magnitude of rotation frequency is relatively high due to the linewidth of the optical mode is relatively large. In comparison to optical mode, the mechanical mode possesses much small linewidth, which inspires us to take advantage of this instinctive property of the mechanical mode.

\section{CONCLUSION \label{conclusion}}

We theoretically propose an effective mechanical $\mathcal{PT}$-symmetric scheme that consists of two optical modes driven by the red-detuning and blue-detuning pump laser respectively and two mechanical modes coupling to corresponding cavity modes and coupling to each other simultaneously. Due to driving the two cavities symmetrically, one can manage either mechanical gain or loss by tuning the system parameter values such as the effective optomechanical coupling strength $G_1$ and the coupling strength between the two mechanical modes $J$. In the absence of the rotation, one can prepare the system in EP with appropriate parameters. In the presence of the perturbation, the eigenvalues of the system are driven apart from the EP and move to the opposite directions. We compare the optical frequency shift with the mechanical frequency response. And it turns out that the mechanical shift more than 50 times to optical response when the rotation frequency is sufficiently small even in the same parameter values. Furthermore, the mechanical modes possess narrower linewidth and  can promote sensitivity of gyroscope towards a new level by taking good advantage of the intrinsic property of mechanical modes. For traditional gyroscopes operating at DP, at which the response is proportional to the perturbation strength, the enhancement factor is 1 no matter what the perturbation strength is. However, the sensitivity of the gyroscopes operating at exceptional point depends on the rotation frequency. The smaller the rotation speed can reach, the higher the sensitivity of the gyroscope can get. As the perturbation strength grows, the enhancement factor approaches the limit $\eta \sim 1$, which indicates the efficiency of EP in detecting weak perturbation such as nanoparticles, temperature variation, and rotation frequency. Our effective mechanical scheme combines the superiority of the narrow linewidth of the mechanical modes and the great enhancement performance of the EP. Moreover, we choose system parameter values with caution under the consideration of experimental feasibility to bridge the theoretical proposal and experimental realization. Our work deepens comprehension of EPs and explores an avenue to applying EP into detecting and sensing in optomechanical systems.

\begin{acknowledgments}

This work is supported by the National Natural Science Foundation of China under Grants No. 61727801; Tsinghua University Initiative Scientific Research Program; The National Key R $\&$ D Program of China (2017YFA0303700); Beijing Advanced Innovation Center for Future Chip (ICFC); The Key R $\&$ D Program of Guangdong province (2018B030325002); The China Postdoctoral Science Foundation under Grant (No. 2019M650620 and No. 2019M660605). 

\end{acknowledgments}

\end{document}